\documentclass[review]{elsarticle}

\usepackage[colorlinks,linkcolor=blue,citecolor=blue]{hyperref}

\journal{Fluid Phase Equilibria}
\usepackage[dvipsnames]{xcolor}

\begin{document}

\begin{frontmatter}

\title{NUCLEATION RATES OF WATER USING ADJUSTED SAFT-0 EOS}

\author{Fawaz Hrahsheh\fnref{myfootnote}}
\address{Higher Colleges of Technology, ETS, MZWC, Abu Dhabi, 58855, UAE.}
\address{Al al-Bayt University, Department of Physics, Faculty of Science, Al-Mafraq, Jordan}
\fntext[myfootnote]{fyh44f@mst.edu}

\begin{abstract}
The SAFT-0 is an equation of state (EOS) that considers the effects of molecular association based on the statistical association fluid theory (SAFT). This EOS recently showed relatively
successful calculations of the phase-equilibrium properties and the classical and nonclassical nucleation rates of methanol. Motivated by methanol results, we use the SAFT-0 EOS for water, in particular within the temperature range of anomalous density behavior below $T_{max}=277.15 K$. To do so, we adjust the effective temperature-dependent segment diameter in terms of the association energy in a way that the SAFT-0 EOS reproduces the water vapor-liquid equilibria and the vapor pressures, particularly in the temperature range where the data of nucleation rates of water are available (220-260 K).
The Gibbsian form of classical nucleation theory (CNT) (known as the P-form) and
nonclassical gradient theory (GT) calculations were carried out using
the SAFT-0 EOS with and without including this adjusted diameter. Calculated rates were compared to the experimental
values of W\"olk and Strey [\textit{J. Phys. Chem.} B 2001, \textbf{105}, 11683-11701].		
In addition to the phase-equilibrium properties, this adjustment improved the nucleation rates from both GT and CNT by factors of 500 and 100, respectively. To explore this further, the GT and experimental rates were analyzed using
Hale's scaled model [\textit{J. Chem. Phys.}, 2005, \textbf{122}, 204509]. This analysis shows that the predictions of GT scale relatively well with those of the experimental data.\\
\end{abstract}


\end{frontmatter}

|

\section{Introduction}

First-order phase transitions play an important role in nature as well as in
many technical applications. Simple examples {of first order phase transition} are condensation\cite{rasmussen_nucleation_1992}, evaporation\cite{macdowell2004}, crystallization\cite{tabazadeh2002, Hallett1964}, and melting\cite{murata2016}. {Such phase transitions have an energy barrier equal to} the work of formation of a small embryo (or nucleus) of the new phase, which emerges from fluctuations within the "supersaturated" mother phase. This initiating process of most first-order phase transitions is called nucleation\cite{feder1966, Frenkel1996}. The hallmark of	such a transition is the discontinuous change of density, for example, the condensation of supersaturated vapor into liquid droplets. The first treatment of the thermodynamics of nucleation is due to Gibbs\cite{gibbs_j._willard_scientific_1961}. Gibbs showed that the reversible work required to form a nucleus of the new phase consists of two terms: a bulk (volumetric) term and a surface term.

 Theoretical analysis of nucleation rates is of great importance in connection with atmospheric aerosol formation and materials synthesis, clustering and condensation in vapors, crystallization of liquid alloys, phase separation in solid solutions, kinetics of colloidal and biological systems and many other growth-related phenomena such as thin film condensation, epitaxy of semiconductor quantum dots and freestanding nanowires\cite{kulmala1996,castleman1981,spracklen2006,curtius2006}. 
 
 One of the simplest examples to illustrate the mechanism of nucleation is the formation of a small liquid droplet in a supersaturated vapor\cite{viisanen1997}. If we compress a vapor at
constant temperature, the condensation will not commence at the saturation pressure,
but above it: the vapor remains in a metastable state for some time until thermal fluctuations
form a sufficiently large cluster or nucleus, which can grow spontaneously.
Central to nucleation theory is the expression for the nucleus (critical-size nanodroplet for spontaneous growing) work of formation. In the 1870s Gibbs~\cite{gibbs_j._willard_scientific_1961} showed that this reversible work equals the difference between the free energy of the metastable phase with and without the droplet present. This macroscopic change of free energy associated with the formation of a nucleus consisting of monomers always contains a volume term and a surface energy term.

Classical nucleation theory originated with the work of Volmer and Weber~\cite{volmer_keimbildung_1926} in 1926. By using the kinetic theory of gases and equilibrium thermodynamics, they derived an expression for the nucleation rates. Farkas~\cite{farkas_keimbildungsgeschwindigkeit_1927} (1927), Becker-D\"oring~\cite{becker_kinetische_1935} (1935 ), Frenkel~\cite{frenkel_general_1939} (1939) and Zeldovich~\cite{j._b._zeldovich_theory_1942} (1942)  established the steady-state version of the so-called classical nucleation theory (CNT). The CNT considers the droplet as a uniform bulk phase separated by a sharp interface from the old phase (metastable vapor). The distinct feature of this theory is that it needs experimentally accessible bulk thermodynamic properties to calculate the nucleation rates.

Calculations of the nucleation rate $J$ are based on the so-called Becker-D\"oring~\cite{becker_kinetische_1935} expression. They assumed that during nucleation, a cluster grows by gaining a molecule at a rate known as the nucleation (or condensation) rate. However, most experiments show that the classically calculated nucleation rates have an incorrect temperature dependence and poor agreement with experimental data~\cite{sinha_argon_2010, wolk_homogeneous_2001, brus_homogeneous_2008}.
Nonclassical nucleation theories using the density functional theory (DFT)~\cite{oxtoby_nonclassical_1988} and the gradient theory (GT)~\cite{wilhelmsen2015,wilhelmsen2014,rowlinson_translation_1979,cahn_free_1959,bedeaux2003,granasy1994,johannessen2004,obeidat_nucleation_2004}, an approximation of DFT, predict a $T$-dependence in good agreement with experiment. Instead of the sharp discontinuity of density in the classical nucleation theory (CNT), DFT and GT treat the droplet as a non-uniform system whose density varies continuously with distance from the center of the droplet, eventually reaching the value of the surrounding mother phase.

Nucleation, as the first stage, determines many important
properties of the newly forming phase such as the number density, and the size distribution
of the nuclei\cite{kozisek2009,kozisekSize2017}. We need to understand the nucleation process in order to control
these parameters in experiments or technical applications. For instance, the number and size of nucleating water droplets has an impact on the efficiency and lifetime of a steam turbine as well as the color and reflectivity of clouds, which in turn influence the greenhouse effect. Nucleation of water is perhaps the most popular example in the literature, due to its relevance to industry; in boilers or turbines, in biological systems,
or in atmospheric sciences and climate modeling\cite{spracklen2006}. A previous calculation applied to water using the cubic perturbed hard body (CPHB) with the $P$-form of CNT improved the predicted nucleation rate values by several orders of magnitudes compared to the conventional CNT, but it failed to improve the predicted temperature dependence of the rates~\cite{obeidat_nucleation_2004}. The results of another calculation applied to methanol using the statistical association fluid theory SAFT-0 EOS~\cite{chapman_new_1990} with the GT and the $P$-form of CNT~\cite{obeidat_gradient_2006,obeidat_nucleation_2010-2}  have motivated us to extend the study to water. The semi-empirical Hale model~\cite{hale_temperature_2005, hale_application_1986, hale_scaling_1992} of nucleation rate shows that the methanol rate data exhibit anomalous $S-T$ dependence while the GT rates using SAFT-0 EOS scale remarkably well, illustrating that the Hale plot can be used to assess the theoretical results and the experimental data, as well.

 {Despite several extensions and theoretical advances, classical nucleation theory (CNT) is still the most widely used theoretical tool used to predict nucleation rates\cite{vehkamaki2006,debenedetti1996,kalikmanov2013,kashchiev2000}. Since the popular classical nucleation theory (CNT) deviates by orders of magnitude from experiments for most substances\cite{iland2007}, Wilhelmsen et. al.\cite{wilhelmsen2015,wilhelmsen2015a} investigated whether part of this discrepancy can be accounted for by the curvature-dependence of the surface tension. To that end, they evaluated the leading order corrections for water, the Tolman length and the rigidity constants, using square gradient theory coupled with the accurate CPA (Cubic-Plus-Association) equation of state\cite{kontogeorgis2006,michelsen2001}. The Helfrich expansion\cite{helfrich1973a} is then used to incorporate the curvature corrections into the CNT-framework. For water, this curvature-corrected CNT corrects the wrong temperature dependence of the nucleation rates given by the traditional CNT. Because the CPA equation of state does not reproduce the experimental water data below the temperature of maximum density\cite{guo2017} ($T_{max}\simeq 277.15K$), they applied a linear extrapolation to obtain the coefficients in the Helfrich expansion for water below this temperature where many of the nucleation experiments for water have been performed.}

Similar to CPA EOS, the SAFT-0 is cubic and accurate equation of state, thus it is convenient to use it with the gradient theory to calculate the nucleation rates of non-polar and polar compounds, especially water. In this paper, we adjust the prerequisites of the SAFT-0 EOS itself in a way that it reproduces the vapor pressures and the anomalous water binodal lines below $T_{max}=277.15 K$ in a good agreement with the experimental data~\cite{wagner_iapws_2002}.
Moreover, we apply the GT and CNT to water using SAFT-0 EOS to see if this adjustment will yield improved rates predictions in the temperature range where most of the nucleation rates were measured~\cite{wolk_homogeneous_2001, miller_homogeneous_1983}. It is imperative to mention that the difference between our work and Wilhelmsen et. al. is that we aim to improve the calculated nucleation rates of water by adjusting the SAFT-0 EOS itself, whereas they do so by incorporating the curvature
dependence of the surface tension in the CNT. Interestingly, our results support their work very well.  {The distinct advantage of adjusting the SAFT-0 EOS itself is the simplicity and flexibility of applying it on more complex systems, particularly the aqueous binary and ternary mixtures below $T_{max}$, using the regular mixing rules. Also, adjusting the temperature dependent segment diameter, only, keeps the predictivity and cubicity of SAFT-0 EOS.}

\section{Theory and Computational Approaches}

The reversible work $W$ needed to create a critical-size cluster of the new phase in a metastable vapor is given by Gibbs~\cite{gibbs_j._willard_scientific_1961} as:
\begin{equation}
W=A\gamma-V(P_l-P_v).
\label{WofForm}
\end{equation}
Here, $A$ and $\gamma$ are the surface area and surface tension, respectively, of the cluster, V is its volume, $P_l$ and $P_v$ are the internal pressure of the new bulk reference phase and the actual pressure of the mother phase, respectively, at the same value of the chemical potential for both phases. Using the Laplace equation for the pressure difference between two phases separated by a curved surface, Gibbs~\cite{gibbs_j._willard_scientific_1961} found an explicit relation for the work of formation $W$ of critical size droplet as:
\begin{equation}
W=\frac{16\pi}{3}\frac{\gamma_\infty^3}{(P_l-P_v)^2}.
\label{Pform}
\end{equation}
Here, the curved surface tension $\gamma$ is approximated by the experimentally reachable values of flat interfaces $\gamma_\infty$ because of a lack of knowledge of the exact surface tension $\gamma$ for typical critical droplets (one nanometer or less radii). Obeidat et al.~\cite{obeidat_gradient_2006,obeidat_nucleation_2004} called Eqn.\ref{Pform} the $P$-form of work of formation. Gibbs' method for calculating the pressure $P_l$ will be described below.

Calculating the accurate work of formation of the critical-size droplets is the challenge to use the Becker-D\"oring~\cite{becker_kinetische_1935} expression of nucleation rate at a specific temperature $T$,

\begin{equation}
J=J_0\exp{\left(-\frac{W}{k_BT}\right)},
\label{BD_NucRate}
\end{equation}
where $k_B$ is Boltzmann constant, and the pre-exponential factor $J_0$ is given as~\cite{obeidat_nucleation_2004}:
\begin{equation}
J_0=\sqrt{\frac{2\gamma_\infty}{\pi m_v}}{v_l}\left(\frac{P_v}{k_B T}\right)^2,
\label{PreExpJ0}
\end{equation}
where $m_v$ is the mass of a condensible vapor molecule, and $v_l$ is the molecular volume of the new phase.

The Gibbs' expression of the reversible work (Eqn.~\ref{WofForm}) contains two terms: a bulk or volumetric term that thermodynamically stabilizes the droplet, and a surface term that is destabilizing because of the increase of free energy associated with forming new surface. Since the surface term is crucial for using Becker-D\"oring model of nucleation, Cahn and Hilliard~\cite{cahn_free_1959} wrote the Helmholtz free energy $F$ as a functional of the system density $\rho$ with the square-gradient approximation, first proposed by van der Waals~\cite{rowlinson_translation_1979}, to account for the inhomogeneous interfacial region separating the critical droplet from the uniform mother phase. Their functional is given by:
\begin{equation}
F[\rho(r)]=\int{dr f[\rho(r)]}=\int{dr \left(f_0[\rho(r)]+(c/2)[\nabla \rho(r)]^2\right)}.
\label{FGT}
\end{equation}
Here, $f$ and $f_0$ are the Helmholtz free energy densities of the inhomogeneous and homogeneous fluids, respectively, and $c$ is the so-called influence parameter~\cite{li_temperature_2003}, which is evaluated as a function of temperature by forcing agreement between calculated and experimental values of the bulk surface tension. The equilibrium droplet density profile that makes the work of formation an extreme value can be found by solving the following Euler-Lagrange equation:
\begin{equation}
\mu=\mu_0-c\nabla^2\rho(r)
\label{ELEqn}
\end{equation}
where $\mu$ and $\mu_0(\rho)$ are the chemical potentials of the bulk vapor phase and the homogeneous fluid, respectively, at density $\rho$. Expressions for $f_0$ and $\mu_0$ are readily found from a given equation of state. 

Cahn and Hilliard first found the reversible work of critical nucleus formation can be evaluated as:
\begin{equation}
W=\int{[\Delta W+(c/2) (\nabla \rho)^2 ]dV},
\label{WG}
\end{equation}
where $\Delta W=W(\rho)-W(\rho_v)$, $W(\rho)=f_0-\rho \mu$, and $\rho$ is the density of the bulk phase. This work of formation may be used in Equation (\ref{BD_NucRate}) to determine the nucleation rates. Obeidat, Li and Wilemski\cite{obeidat_nucleation_2004, obeidat_nucleation_2010-2,li_temperature_2003,aobeidat2003} provide a comprehensive explanation of implementing the GT to calculate the work of formation of critical nanodroplets. In this work we exactly follow their steps but with the adjusted SAFT-0 EOS.

To calculate the binodal points, we derive the functional pressure and chemical potential from the Helmholtz free energy $F$ as:

 \begin{equation}
P(\rho)=\rho^2\left(\frac{\partial F}{\partial \rho}\right)_{T}
 \label{Pressure}
 \end{equation}

\begin{equation}
\mu(\rho)=\rho\left(\frac{\partial F}{\partial \rho}\right)_{T}+F(\rho).
\label{ChemPot}
\end{equation}
Then we determine the coexisting densities of bulk liquid $\rho_{le}$ and vapor $\rho_{ve}$ by solving the simultaneous equations

\begin{equation}
P(T,\rho_{le})=P(T,\rho_{ve}),
\label{PressureEq}
\end{equation} 

\begin{equation}
\mu(T,\rho_{le})=\mu(T,\rho_{ve}),
\label{ChemPotEq}
\end{equation}
 where $\rho_{le}$ and $\rho_{ve}$ are the equilibrium vapor and liquid densities, respectively. Gibbs' assumed the temperature and chemical potential are the same everywhere in the nonuniform system\cite{gibbs_j._willard_scientific_1961}, i.e., $\mu(T,\rho_v)=\mu(T,\rho_l)$.  After subtracting the equilibrium value of chemical potential from both sides of this equation, we obtain
  \begin{equation}
  \mu(T,\rho_{l})-\mu(T,\rho_{le})=\mu(T,\rho_{v})-\mu(T,\rho_{ve}).
  \label{GibbsRef}
  \end{equation}  
Once $\rho_l$ has been found by solving Eq.\ref{GibbsRef}, the reference pressure $P(T,\rho_l)$ (pressure inside the droplet) is straightforward to calculate from EOS. 
	
\section{Adjusted SAFT EOS}	
 The application of perturbation theory to associating systems in
an equation of state was not practical until Wertheim developed his multi-density statistical mechanics
for associating fluids\cite{wertheim_fluids_1984}. Perturbation theories based on Wertheim’s multi-density statistical mechanics have come to be called thermodynamic perturbation theory (TPT). In TPT, there is two energy scales: a short ranged highly
directional association energy scale (hydrogen bonding energy scale) and an orientationally
averaged (non- association) energy scale (reference energy scale). Typically, the reference energy scale accounts for dispersion attractions, dipolar attractions not accounted for
through the association term and higher order multi-pole contributions. Thermodynamic perturbation theory (TPT) forms the basis of the statistical associating fluid theory (SAFT) family of equations of state\cite{chapman_new_1990,gil-villegas1997,gross2002,llovell2006}. Chapman, et al.~\cite{chapman_new_1990} proposed the first equation of state based on the statistical associating fluid theory (SAFT). They wrote the Helmholtz molar free energy $F$ as:
\begin{equation}
F=F^{id}+F^{seg}+F^{chain}+F^{assoc}
\label{SAFT_CHAP}
\end{equation}
where $F^{id}$, $F^{seg}$, $F^{chain}$and $F^{assoc}$ are the ideal term, the segment term, the chain term and the association term, respectively. It is now referred to as SAFT-0~\cite{tan_recent_2008} and we briefly summarized it in a previous work (\cite{obeidat_nucleation_2010-2}). While traditional cubic EOSs such as PR and SRK only have a single energy scale of attraction, SAFT allows for separate accounting of hydrogen bonding and reference (non-hydrogen bonding) attraction degrees of
freedom. The strength of SAFT-0 EOS emerges from the association term $F^{assoc}$ for self-associating compounds which is given by:
\begin{equation}
\frac{F^{assoc}}{RT}=\sum_{A=A,B,...}{\left(\ln X^A-\frac{X^A}{2}\right)}+\frac{1}{2M}
\label{ASSOC_SAFT}
\end{equation}
where $R$ is the universal gas constant, $M$ is the total number of association sites on each molecule, $X^A$ is the mole fraction of molecules not bonded at site $A$, and $\sum_A$ represents a sum over all associating sites on the molecule. { Example for molecules with two attractive
	sites $A$ and $B$ (two-site model will be used below) is given as:}  
\begin{equation}
\frac{F^{assoc}}{RT}={\left(\ln X^A-\frac{X^A}{2}\right)+\left(\ln X^B-\frac{X^B}{2}\right)}+\frac{1}{2}.
\label{ASSOC_SAFT2B}
\end{equation}
The mole fraction of molecules not bonded at sites $A$ and $B$ can, respectively, be calculated as follows:
\begin{equation}
X^A=\left[1+N_{\textnormal{Av}}\sum_{B=A,B}{\rho X^B\Delta^{AB}}\right]^{-1}
\label{MolFrac_SAFT}
\end{equation}
\begin{equation}
X^B=\left[1+N_{\textnormal{Av}}\sum_{A=A,B}{\rho X^A\Delta^{BA}}\right]^{-1}
\label{MolFrac_SAFT}
\end{equation}
where $N_{\textnormal{Av}}$ is the Avogadro's number, and $\rho$ is the molar density of molecules, and $\Delta^{AB} (\Delta^{BA}$) is the association strength, given as:

\begin{equation}
\Delta^{AB}=d^3\left[\frac{2-\eta}{2(1-\eta)^3}\right]\kappa^{AB}\left[\exp{\left(\frac{\epsilon^{AB}}{k_BT}-1\right)}\right].
\label{AssStreng_SAFT}
\end{equation}
Here, $d$ is the effective temperature-dependent segment diameter, $\kappa^{AB}$ is the dimensionless association volume, $\epsilon^{AB}$ is the association energy, and $\eta=\frac{\pi N_{{Av}}}{6}\rho d^3 m$ is the segment packing fraction.

{ The association sites in water can be represented by three different models: four-site model ($M=4$), three-site model ($M=3$), and two-site model ($M=2$). Suresh et. al.~\cite{suresh_multiphase_1992} evaluated these models by Wertheim's thermodynamic perturbation theory (TPT)~\cite{wertheim_fluids_1984, wertheim_fluids_1984-1}. They showed that any of the models may be accurately applied but they recommended the two-site model because its accuracy is at least equivalent to that of the other models and it is more convenient to apply in general. Taking into account the results of Wertheim's TPT and the recommendation of Suresh et. al., Gross and Sadowski used the two-site model to apply the PC-SAFT EOS on water\cite{gross2002}. Here, we keep with this choice and the two-site model of water ($M=2$) will be used in this work.  } 

The temperature-independent hard sphere diameter $\sigma$ was related, following the Barcker-Henderson theory~\cite{barker_perturbation_1967}, to an effective temperature-dependent segment diameter, $d(T_r,m)$ as~\cite{chapman_new_1990}:
\begin{equation}
d\left(T_r,m\right)=\sigma f\left(T_r,m\right).
\label{d_BH}
\end{equation}
Here, $f\left(T_r,m\right)$ is a generic function of the reduced temperature ($T_r=\frac{K_BT}{\varepsilon}$) by the Lennard-Jones intermolecular energy parameter $\varepsilon$, $m$ is the number of segments per molecule, and $\sigma$ is the temperature-independent hard sphere diameter. Barker and Henderson\cite{barker_perturbation_1967} introduced a temperature dependent form for $d$ as
	\begin{equation}
	d=\sigma\int_0^1{\left[1-\exp{\left(\frac{-u(z)}{k_BT}\right)}\right]}dz,
	\label{d_integral}
	\end{equation}
 where $u(z)$ is the Lennard-Joens potential and $z$ is the center-to-center distance between interacting (nonbonded) segments. Cotterman et. al.~\cite{Cotterman86} numerically solved this integral for several temperatures and they determined an empirical formula for $d$ by fitting its values for the spherical Lennard-Jones molecules as a function of reduced temperature $T_r$:
\begin{equation}
d\left(T_r,m\right)=\sigma \left[\frac{1+0.2977T_r}{1+0.33163T_r+f(m)T_r^2}\right],
\label{d_SAFT}
\end{equation} 
where $f(m)=0.0010477+0.025337(m-1)/m$ is just an abbreviation in terms of $m$. For SAFT-0 EOS, Chapman et. el.~\cite{chapman_new_1990} used the same empirical formula obtained by Cotterman et. al. for the Lennard-Jones potential. Although $d\left(T_r,m\right)$ does not heavily change with $T$, it still can have a strong effect on the thermodynamics of the system, hence on the curves of the liquid-vapor coexistence.

 It is known that water is highly polarized in the liquid
state. Through detailed first principles
quantum mechanics calculations for small water clusters\cite{ludwig2001}, it has been shown that water hydrogen bond energy,
polarization, and dipole moment\cite{gregory1997,kemp2008} depend on the number
of times the water molecule is hydrogen bonded, as well
as the type of hydrogen bonded cluster. Hence, there are both
hydrogen bond cooperativity and cooperativity between the
hydrogen bonding and the non-hydrogen bonding (reference)
energy scales. Logically, this means that the reference energy
in TPT should depend on the degree of hydrogen bonding\cite{marshall2016,marshall2017}. Marshall\cite{marshall2016} have proposed a methodology to couple the reference energy scale to the degree of hydrogen bonding in the fluid with $d=\sigma$. Applying this methodology on water gave improved predictions of water-hydrocarbon mutual solubilities above $T_{max}$, but it did not improve the predictions of the pure water density or pressure. Later, Marshall\cite{marshall2017} developed a second order thermodynamics perturbation theory (TPT2) to include hydrogen bond cooperativity for the case where a water molecule is a donor and a acceptor, at the same time.  Also, as is common in perturbation theories, TPT2 with PC-SAFT and $d=\sigma$ did not reproduce the density maximum of water. Replacing $\sigma$ with $d(T)<\sigma$ in the hard sphere contributions has the effect of increasing density as compared to the case where the hard sphere diameter $\sigma$ was used in these contributions. Therfore, Marshall described the reason of using $d=\sigma$ for water as following: Using $d(T)<\sigma$ increases the average number of hydrogen bonds per water molecules with a decrease
in temperature which results in the shortening of the hydrogen
bonds. Marshall attributed the failure of predicting the water density maximum to that the density maximum is a result of the change in the water structure (hydrogen bonds number and length), while TPT2 addressed only energetics.

 Recently, Held et al.\cite{held2008}, Cameretti et al.\cite{cameretti2005}, Cameretti\cite{cameretti2009} reported that accurate modeling of liquid densities of water cannot be obtained without modification of its temperature-dependent segment diameter. To do so, we take the exponential form of the temperature-dependent effective diameter in Chen and Kreglewski work~\cite{Chen77} where the Barker-Henderson integral equation~\cite{Barker67} was solved using a square-well potential as a suggestion to use exponential form for our desired adjustment. This exponential form is also supported by Held et al.\cite{held2008} where they added a temperature-dependent exponential term to the temperature-independent hard sphere diameter $\sigma$ in ePC-SAFT EOS\cite{cameretti2005} to treat the deviation of calculated water density between $273.15K$ and $373.15K$. Later, Pereda et al.\cite{pereda2009} studied the temperature dependence of the hydrocarbon solubility in water using the group contribution with association equation of state (GCA-EOS)\cite{zabaloy1993,gros1996}. As a result of their study, they concluded that the temperature-dependent effective diameter $d$ of water has a strong influence on the temperature dependence of the hydrocarbon solubility in water. Therefore, they introduced an exponential correction to $d$ to improve the predictions of water-hydrocarbons mutual solubility within the temperature range $298-353K$ of their experimental data.

 In the present work, we aim to couple the additive hydrogen bonding energy scale and the reference energy scale by including the association energy $\epsilon^{AB}$ in the calculations of the effective temperature-dependent segment diameter. Then, we just need to find a way of adjusting $d$ in terms of $\epsilon^{AB}$ and $T$ so that the SAFT-0 EOS successfully reproduces the water densities and the vapor pressures, particularly for $T<T_{max}=277.15K$. Because the SAFT-0 EOS with the Cotterman's form of $d$ successfully predicts the water binodal line for $T>T_{max}$ (with small deviations in the vicinity of $T_{max}$) whereas it over-predicts it below $T_{max}$, we propose two conditions for the desired adjustment: It should quickly vanish above $T_{max}$ and it should quickly increase below it. To apply these conditions in adjusting $d$ with the site-site association energy, we introduce an exponential correction in terms of the dimensionless distance of the temperature from the water maximum density temperature $\frac{T-T_{max}}{T_{max}}$ and the scaled association energy $\frac{\epsilon^{AB}}{k_B T_{max}}$. This correction is exclusive for water and the adjusted temperature-dependent segment diameter in SAFT-0 EOS becomes
\begin{equation}
d'\left(T_r,m\right)=d\left(T_r,m\right)+\sigma\lambda \exp\left[-\frac{2}{3}\frac{\epsilon^{AB}}{k_B T_{max}}\left(\frac{T-T_{max}}{T_{max}}\right)\right].
\label{dprime_SAFT}
\end{equation}
Here, $\lambda\sigma=0.02355$ is the correction at $T_{max}$ where we calculate it by iteratively adding small corrections to $d$ until we successfully reach the corresponding water experimental density (the maximum density). The liquid water's anomalous behavior due to hydrogen bonding effect is accounted for by employing $\frac{T-T_{max}}{T_{max}}$ to make the adjustment solely influential below $T_{max}$. The adjusted temperature-dependent segment diameter is bounded by $T=203\pm 5K$ where a minimum density is evident there\cite{Mallamace18387}. As far as we know, adjusting the temperature-dependent hard sphere diameter have never been made by including the association energy.  We already tested this adjustment for four-site ($M=4$) model of water in a different subject of study that will be published in the near future.\\

In this paper, we apply the CNT~(\cite{obeidat_nucleation_2010-2, obeidat_nucleation_2004}) and GT~(\cite{li_temperature_2003, obeidat_nucleation_2010-2}) to calculate the nucleation rates of water using the adjusted SAFT-0 EOS. Its parameters for water (given in Table~\ref{SAFT:PARAM}) are obtained by simultaneously fitting the experimental saturated vapor pressures and liquid densities~\cite{wagner_iapws_2002}.

\begin{table}[tb!]
	\caption{Fitting parameters of SAFT-0 EOS for two-site water.}
	\renewcommand{\arraystretch}{1.2}
	\begin{tabular*}{85mm}{|c@{\extracolsep\fill}cccc|}
		
		\hline
		$\sigma$ [\AA]  & $\varepsilon/k_B [K]$ &$m$&$\epsilon^{AB}/k_B$ [K]& $\kappa^{AB}$   \\
		\hline
		2.925  & 294.1 &1.026&2938.8& 0.053   \\
		\hline
	\end{tabular*}
	
	\label{SAFT:PARAM}
\end{table}

\section{Results and Discussions}

\begin{figure}
	\includegraphics[width=8cm]{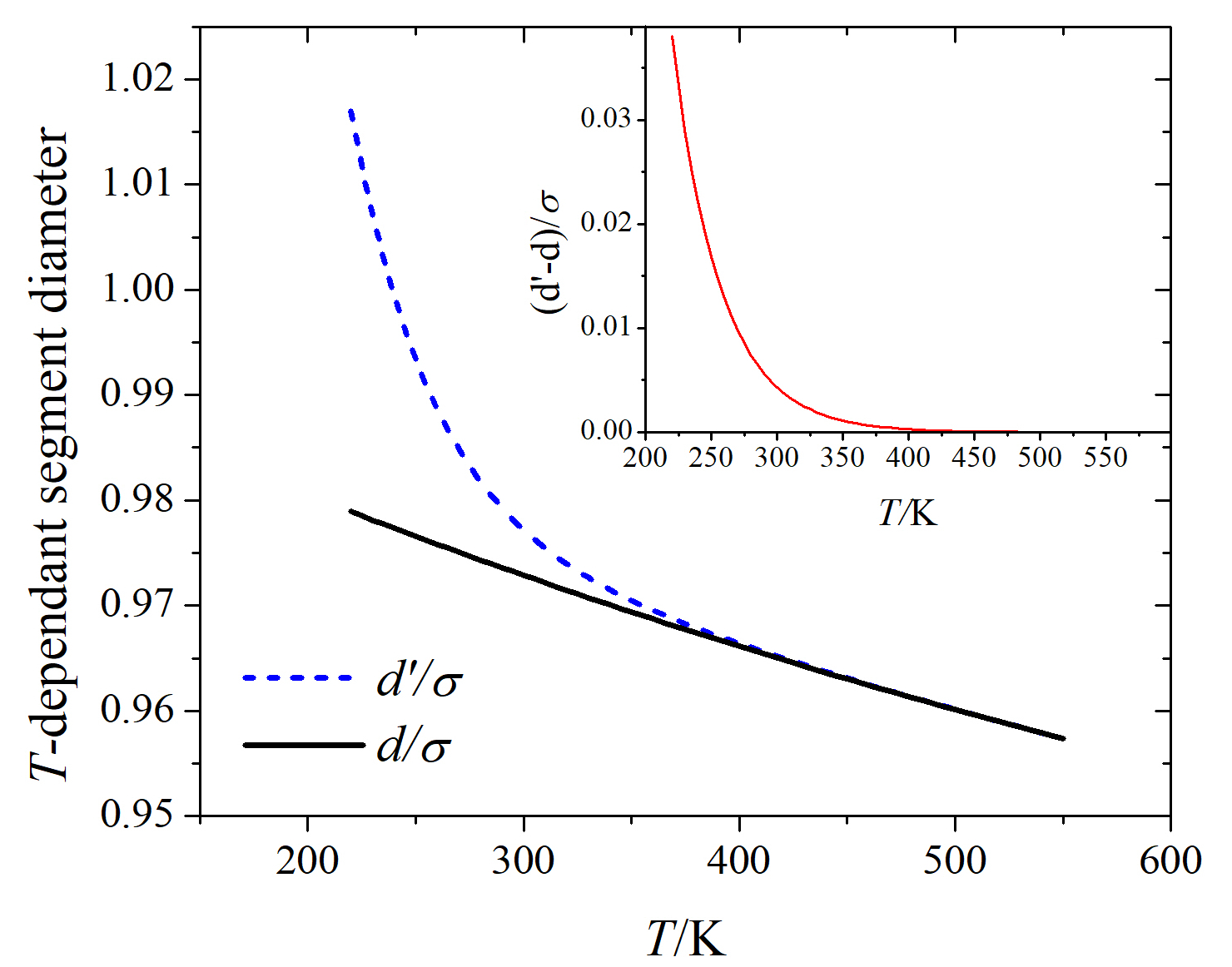}
	\caption{ Comparison between the temperature-dependent segment diameter $d$ (solid line) and the adjusted one $d'$ (dashed line). Inset: The differences between these two diameters represent the adjustment term vs. temperature. }
	\label{fig:dvsT}
\end{figure}

The effect of the adjustment on the temperature-dependent segment diameter is shown in Figure~\ref{fig:dvsT}.  {The fast increase of $d'$, compared to $d$, as $T$ drops down is the distinct feature of this adjustment. Specifically, this exponential increase of $d'$ enables the SAFT-0 EOS to calculate the anomalous behavior of the supercooled water density below $T_{max}$ where the larger adjustment (as shown in the inset), the larger decrease in the liquid water density.}\\

\begin{figure}
	\includegraphics[width=8cm]{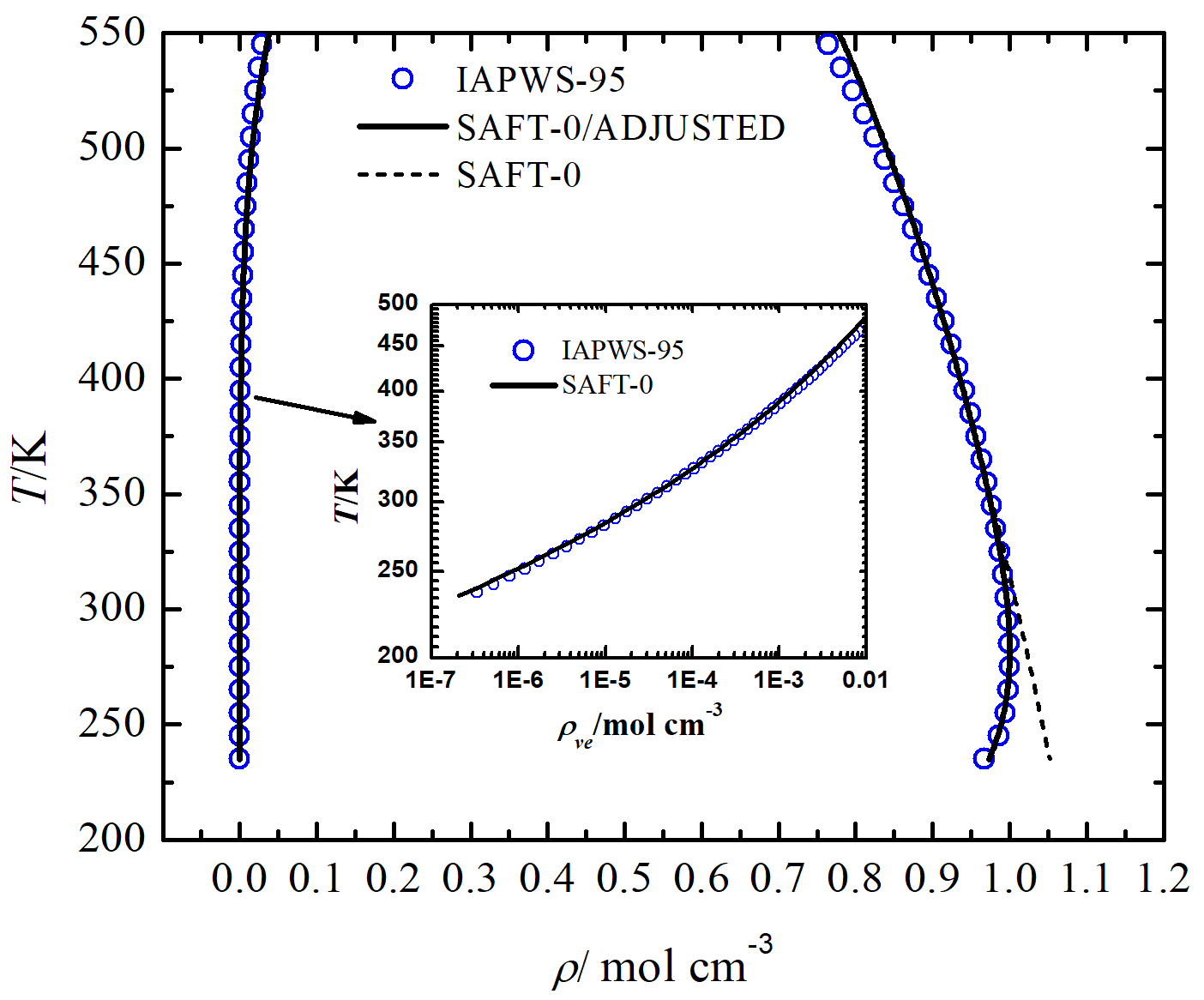}
	\caption{Binodal densities of water calculated using SAFT-0 and SAFT-0/ADJUSTED compared with IAPWS-95 values~\cite{wagner_iapws_2002}. Inset: Log-Plot of saturated vapor density vs. T.}
	\label{fig:binodal}
\end{figure}

 The predictions of SAFT-0 for the equilibrium vapor-liquid densities of water compared to the values calculated with the IAPWS-95 formulation~\cite{wagner_iapws_2002} are shown in Figure~\ref{fig:binodal} for several temperatures. IAPWS-95 is an analytical equation based on a multiparameter fit of all the experimental data available at temperatures above $234 K$~\cite{wagner_iapws_2002,obeidat_nucleation_2004}. Note that the results of the IAPWS-95 EOS may be regarded as the experimental values since this equation accurately treats the anomalous compressibility of supercooled liquid water and describes it to high precision above $234K$. Starting with the results of regular SAFT-0 EOS, we see that this equation accurately predicts the equilibrium vapor-liquid densities for $T>335 K$, but it is severely deficient in predicting the anomalous liquid binodal line below $T_{max}$ (dashed line). Later, we will see the effect of this mis-prediction on calculating the nucleation rates of water. The adjusted SAFT-0 EOS fairly resolves this inaccuracy and predicts the binodal points (solid lines) within the range of temperature $220K-445K$ very well with minor deviations near the range boundaries. However, one may reduce this deviation by calculating the fitting parameters of SAFT-0 EOS and the new parameter ($\lambda$) within a shorter temperature range such as $210K-300K$. In present work, we provide the acceptable fitting parameters in a very wide temperature range to make the adjusted SAFT-0 EOS ready for other studies. \\

\begin{figure}
	\includegraphics[width=8.5cm]{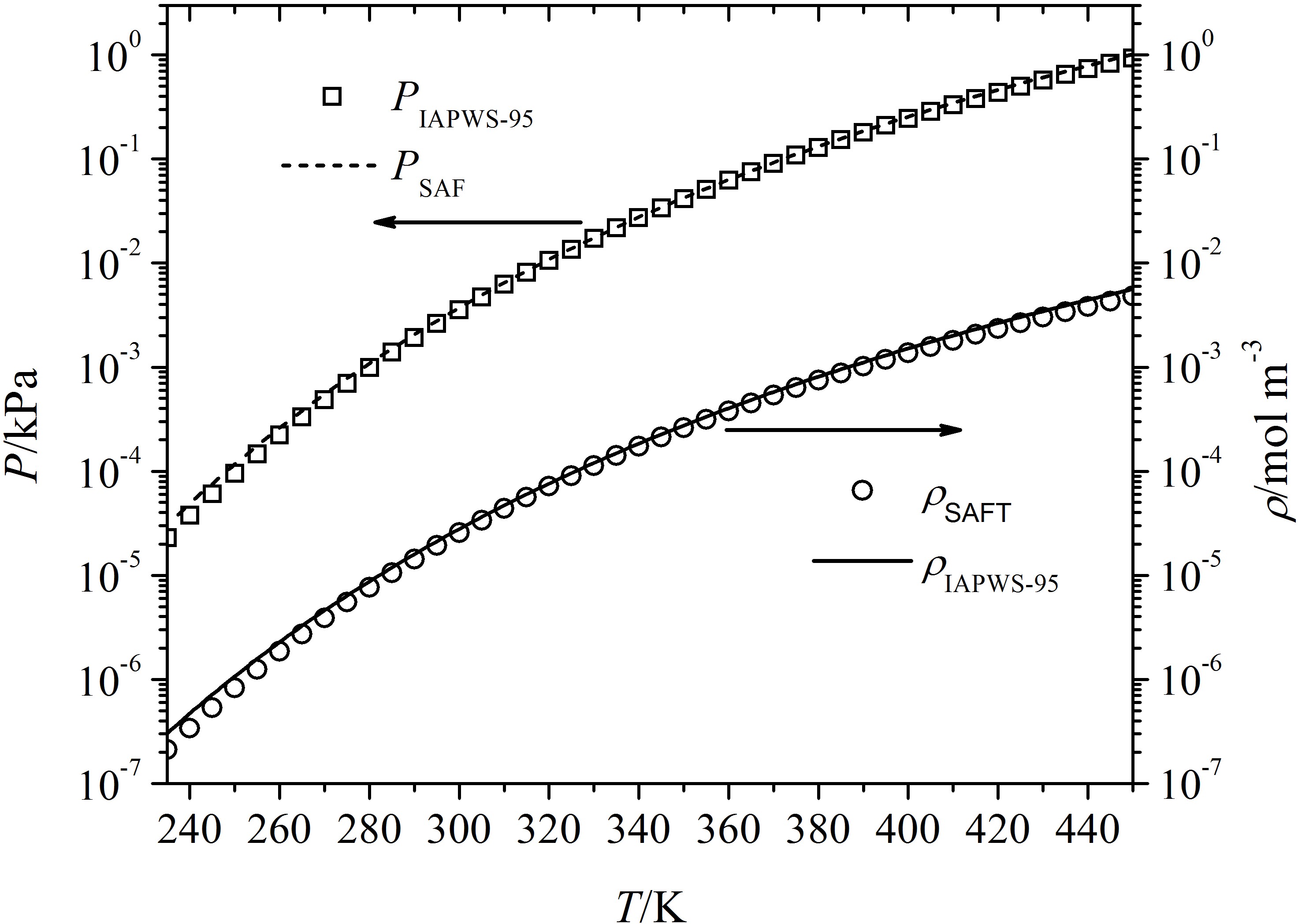}
	\caption{Log. plot of saturated vapor pressure $P$ and density $\rho$ of water calculated using $T$-adjusted SAFT-0 EOS compared with  IAPWS-95 values~\cite{wagner_iapws_2002}. }
	\label{fig:Pv_Nv}
\end{figure}

Figure~\ref{fig:Pv_Nv} compares accepted~\cite{wagner_iapws_2002} binodal vapor densities and equilibrium vapor pressure of water with values calculated using the adjusted SAFT-0 EOS. We see the calculated results agree very well with experimental data in the temperature range $235- 450K$, in particular the range relevant for nucleation measurements. The small deviation at very low temperature is due to the same reason mentioned above.\\

\begin{figure}
	\includegraphics[width=8cm]{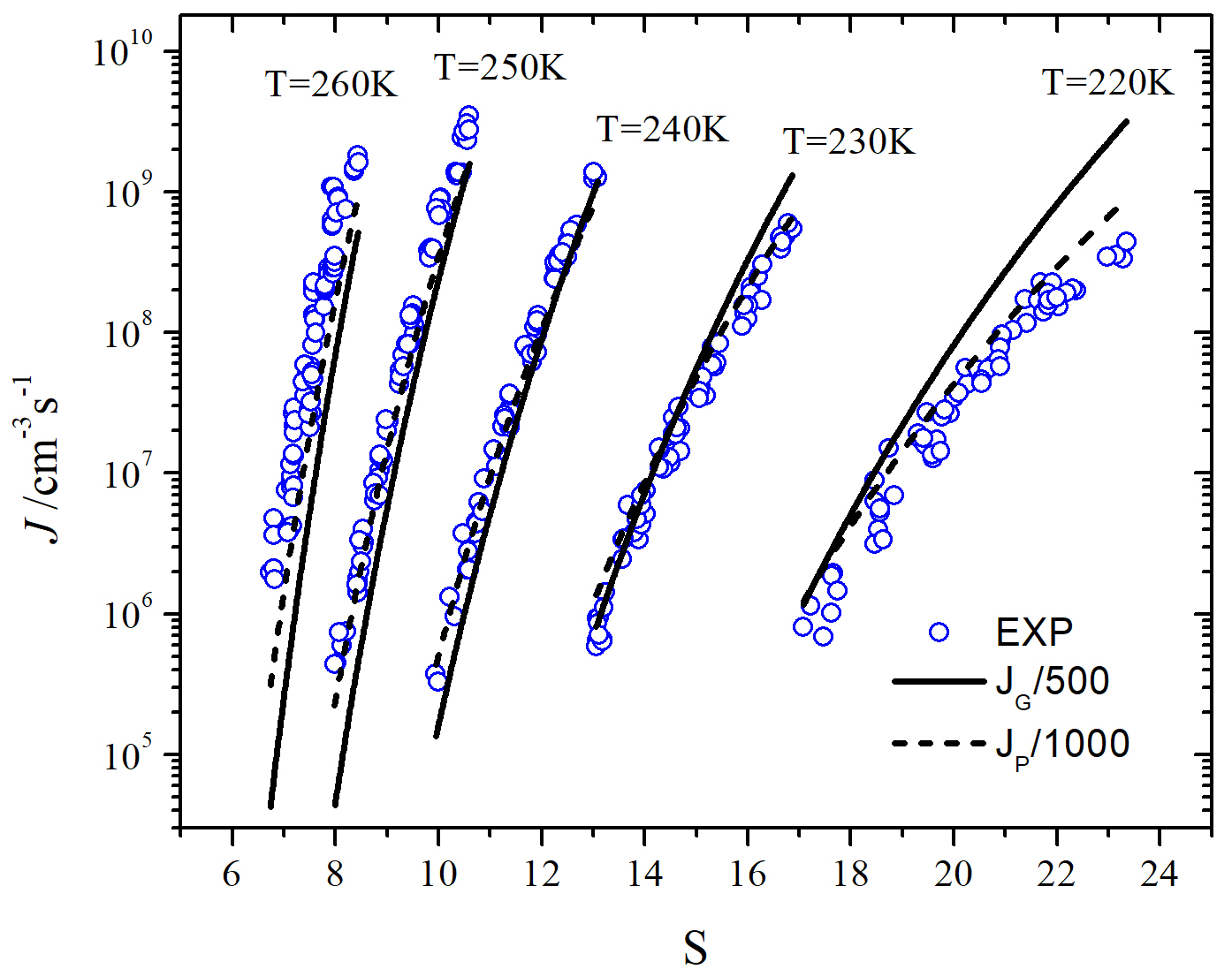}
	\caption{ Nucleation rates of water using the SAFT-0 EOS with CNT and GT compared with experimental rates~\cite{wolk_homogeneous_2001}. $J_G$: gradient theory; $J_P$: $P$-form of CNT. }
	\label{fig:NucSaft}
\end{figure}

Both forms of the classical and nonclassical nucleation theories have been used with the SAFT-0 EOS to determine the nucleation rates of water. For comparison, Fig.~\ref{fig:NucSaft} compares the predicted nucleation rates of water using the SAFT-0 EOS in terms of Cotterman's form of $d$ with W{\"o}lk and Strey measurements~\cite{wolk_homogeneous_2001}. {The reason for using W{\"o}lk and Strey measurements will be shown later below}. We note that the GT and $P$-form roughly show a similar dependence on $S$ ($P$-form has a slightly better slope at low temperatures) , but the $P$-form has a better $T$ dependence (the gaps between the experimental and predicted nucleation rates are smaller, particularly for small temperatures) and is about one order of magnitude higher than the GT. Neither GT nor CNT reproduces the $J$ values very well.\\

\begin{figure}
	\includegraphics[width=8cm]{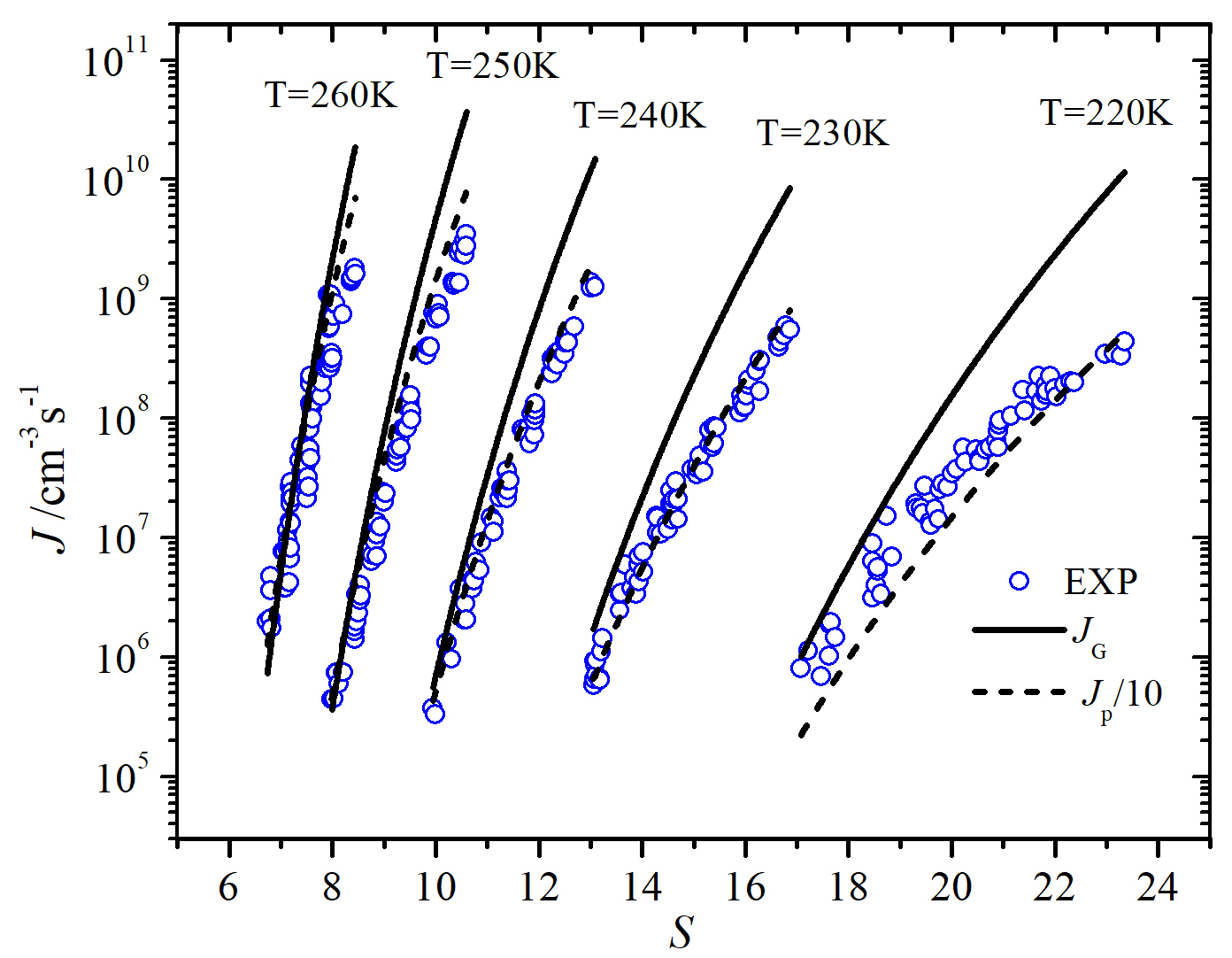}
	\caption{ Nucleation rates of water using the adjusted SAFT-0 EOS with CNT and GT compared with experimental rates~\cite{wolk_homogeneous_2001}. $J_G$: gradient theory; $J_P$: $P$-form of CNT. }
	\label{fig:NucASaft}
\end{figure}

We see in Fig.~\ref{fig:NucASaft} that the nucleation rates using adjusted SAFT-0 EOS are considerably improved compared to the results in Fig.~\ref{fig:NucSaft}. First, the magnitudes of the nucleation rates are improved by factors of 500 and 100 for GT and CNT, respectively, but the P-form of CNT shows a slightly better $S$ dependence. Second, we see that adjusting $d$ by including the association energy significantly improves the $T$ dependence of GT. 
Both the CNT and GT rates share the following features. At low temperatures in the figure \ref{fig:NucASaft}, one sees poorer agreement with the data than at the higher temperatures because of the use of a constant scale factor. 

Next, we apply a $T$-dependent scaling factor and we discuss the $T$ dependence of the rates in more detail. Since the GT calculations of nucleation rates of water are better by a factor of 10 than those from the P-form of CNT and because the GT usually gives better results for other associating compounds than the CNT\cite{obeidat_nucleation_2010-2}, its predictions of nucleation rates will be assessed by Hale's $T$-dependent scaling models~\cite{hale_temperature_2005,hale_scaling_2004,hale_scaling_1992,hale_application_1986}. Hale's scaled nucleation rate equation is, perhaps, the most useful semi-empirical model that accurately describes the temperature dependence for many simple vapor systems. It can be used to examine the accuracy of nucleation rate data. For example, Obeidat et al.~\cite{obeidat_nucleation_2010-2} used Hale plot to assess Strey et al.~\cite{strey_homogeneous_1986,strey_effect_1986} measurements of methanol nucleation rates. Hale plots showed that the methanol rate data exhibit anomalous $S-T$ dependence. Obeidat et al. attributed this to the inadequacy of the thermodynamic data base used by Strey et al. to correct the original $S-T$ data for the effects of gas phase association.

\begin{figure}
	\includegraphics[width=8cm]{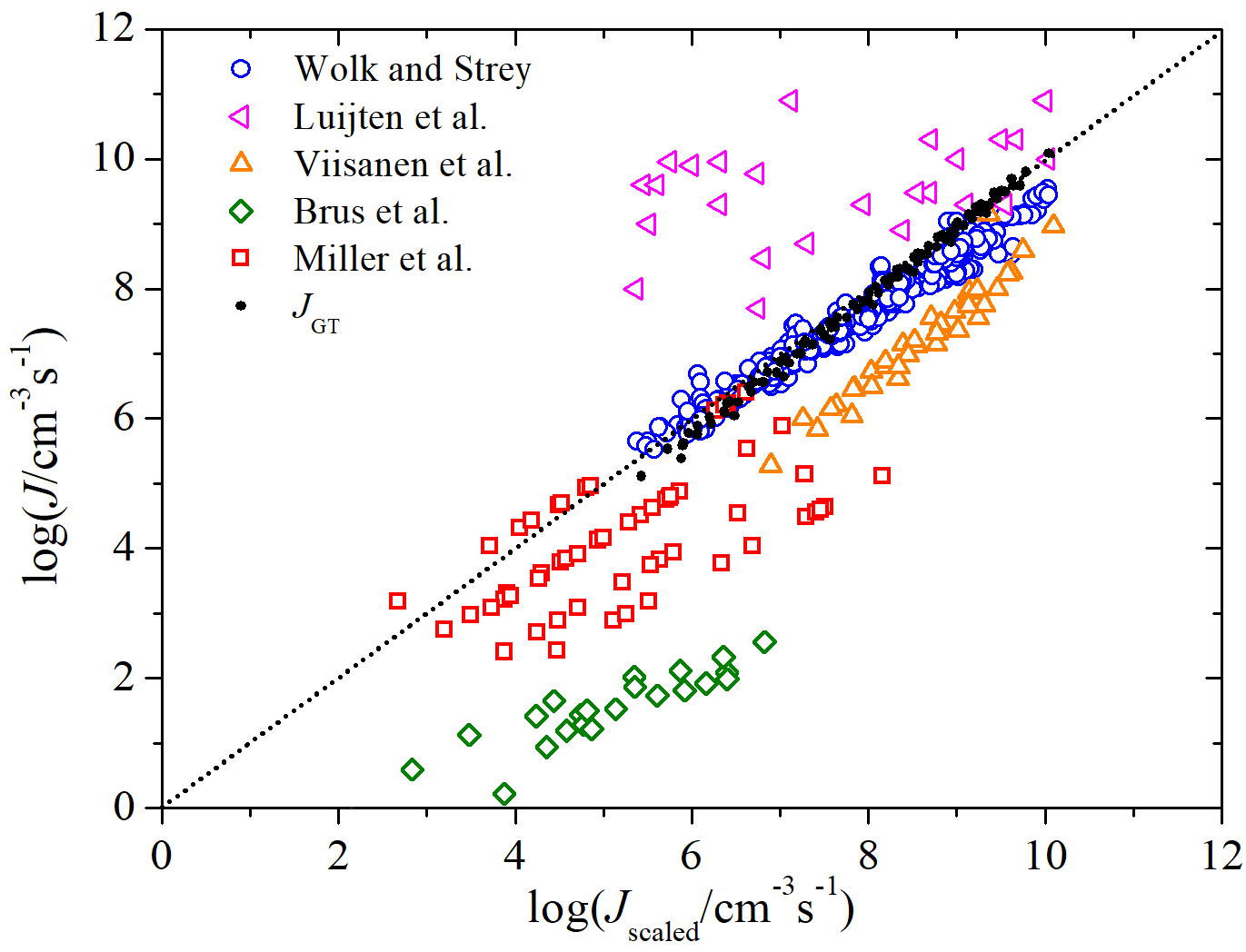}
	\caption{ A comparison of the homogeneous nucleation rates of water predicted by Hale's scaled model with several experimental data sets and our GT results, with $\Omega=1.45$.	The data sets are from the following references: W{\" o}lk and Strey\cite{wolk_homogeneous_2001}, Luijtenet al.\cite{luijten1997}, Viisanen et al.\cite{viisanen1993}, Brus et al.\cite{brus2008a}, and Miller et al.\cite{miller_homogeneous_1983}. The dotted line indicates perfect agreement. }
	\label{fig:JexpvsJscl}
\end{figure}

{Hale defined a simple scaled model of nucleation rate by exploiting scaled expressions for the vapor pressure and for the surface tensions\cite{hale_scaling_2004,Hale1992} as follows:
	\begin{equation}
	J_{scaled}=J_0 \exp{\left[ -\frac{16\pi\Omega^3\left(\frac{T_C}{T}-1\right)^3}{3\left(\ln S\right)^2} \right] },
	\label{Jscaled}
	\end{equation}
	where $J_0$ is the kinetic prefactor for the steady state homogeneous nucleation rate, $T_C=647.15K$ is the experimental critical temperature of water, and $\Omega=\sigma_0/k/\rho^{2/3}$ is the excess surface entropy per molecule estimated from the experimental values of surface tension (approximately 2 for normal liquids and 1.5 for polar substances). Here, $\sigma_0$ is a material-dependent constant obtained by fitting the surface tension data as a linear equation of surface tension $\gamma=\sigma_0(T_C-T)$, $k$ is the Eotvos constant\cite{MacDougall1936,hale_application_1986}, and $\rho$ is the liquid number density. A comparison of the scaled model of nucleation rate with some experiments and GT is shown in Figure~\ref{fig:JexpvsJscl}. Here, the experimental data and the GT predictions are plotted versus $J_{scaled}(T_{exp},S_{exp})$, with $\Omega=1.45$. The latter value of $\Omega$ is in the expected range for polar substances (calculated by Hale et al.\cite{hale_scaling_2004}). In addition to our calculated values of nucleation rates of water using GT, this comparison shows that W{\"ok} and Strey measurements scale very well and lie along the perfect agreement line.}  

\begin{figure}
	\includegraphics[width=8cm]{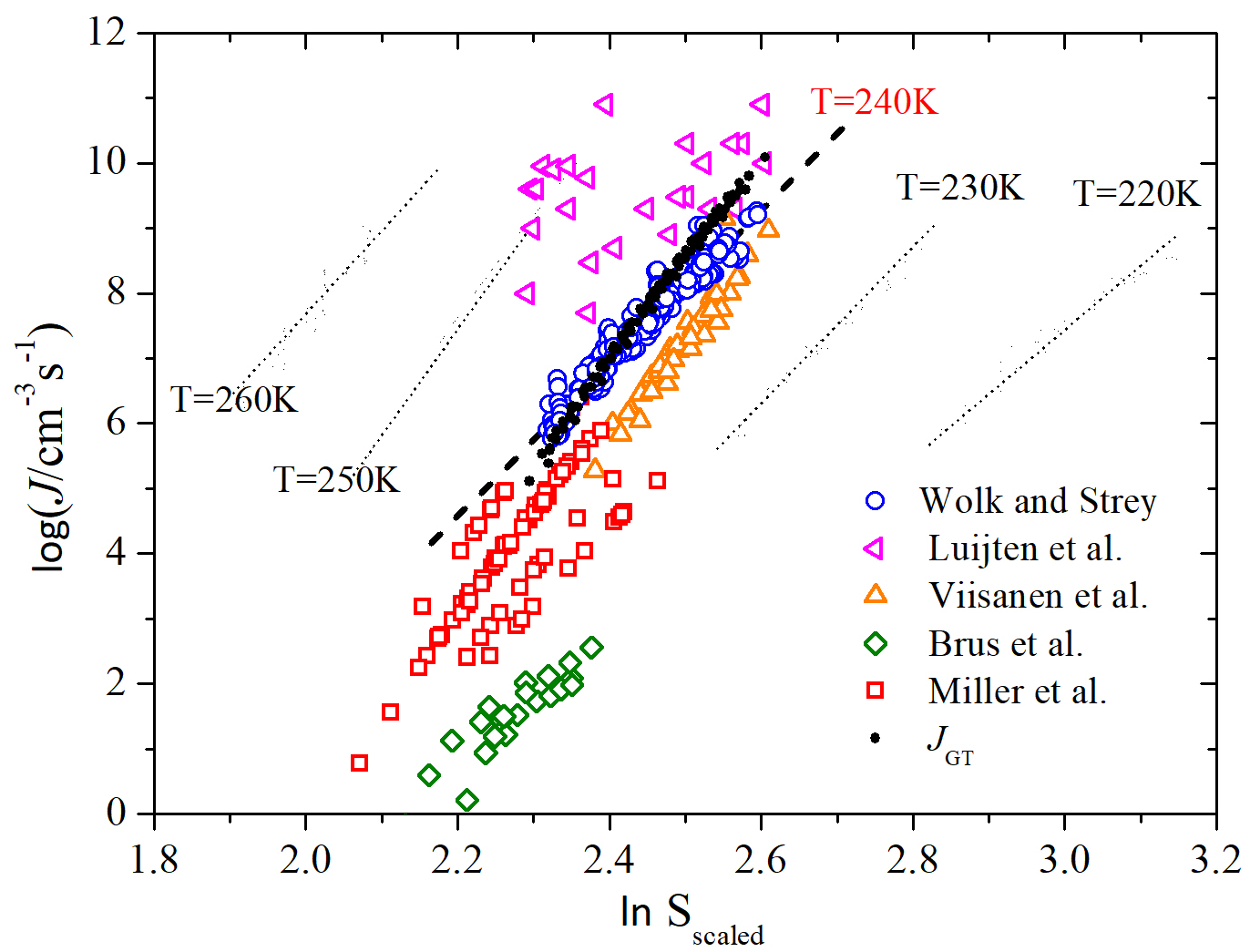}
	\caption{ The GT predictions of homogeneous nucleation rates of water and experimental data plotted versus the scaled supersaturation function, $\ln S/(T_c/T-1)^{3/2}$. The scaled supersaturation is multiplied by a	normalizing constant, $C_0=(T_c/240-1)^{3/2}$, so that the values fall in the same	range of $T=240 K$ (dashed line). {The dotted lines indicate where the W{\"o}lk and Strey's measurements fall if $\log J_{exp}$ is plotted	versus $\ln S$ (before scaling $S$)}. The references for the data are listed in the caption of Fig.\ref{fig:JexpvsJscl}.}
	\label{fig:HaleNuc2}
\end{figure}

Hale's plots provide a simple means of assessing the combined supersaturation and temperature dependence of a set of nucleation rates. For many systems, nucleation rate data from different laboratories often lie on or close to a single, universal line indicating mutual consistency~\cite{gharibeh_homogeneous_2005-1,brus_homogeneous_2006,ghosh_using_2008,ghosh_homogeneous_2010}. An application of the scaling that requires no knowledge of $\Omega$ is shown in Figure~\ref{fig:HaleNuc2} where $\log J$ is plotted versus the scaled supersaturation ($\ln S_{scaled}=\ln S/(T_c/T - 1)^{3/2}$)~\cite{hale_application_1986,hale_scaling_2004}.	
The scaled supersaturation in Figure~\ref{fig:HaleNuc2} is multiplied by a normalizing constant $(T_c/240 - 1)^{3/2}$, so that the values fall in the same range as $\ln S$. Hence, the dashed line at $T=240K$ is the perfect agreement line. We see that the W{\"o}lk and Strey data corresponding to constant temperatures that spread out in the standard $\log J_{exp}$ versus $\ln S$ plot (indicated by the dot lines) collapse onto a single line (indicated by dashed line) when we plot $\log J_{exp}$ versus $\ln S_{scaled}$. Similarly, the plot of  $\log J_{GT}$ versus $\ln S_{scaled}$ shows a good scaling of the nucleation rates of water predicted by applying GT with adjusted SAFT-0 EOS (indicated with filled circles). As expected, the GT results scale very well using Hale's scaling model and show a good temperature-dependent agreement with the W{\"ok} and Strey experimental data, but with slightly stronger dependence on the scaled supersaturation ratio (larger slope).\\

\section{Conclusions}
 In order to solve the inaccuracy of SAFT-0 EOS in predicting the water binodal lines below $T_{max}$, we adjust the temperature-dependent hard sphere diameter by adding an exponential correction in terms of the association energy. The adjusted SAFT-0 EOS predicts the water phase equilibria adequately well within a temperature range of $220K-445K$.

We have made classical and nonclassical nucleation rate calculations for water vapor using the SAFT-0 and adjusted SAFT-0 EOS. The calculated nucleation rates by GT and CNT are compared with W\"olk and Strey measurements~\cite{wolk_homogeneous_2001} using a standard format.  The adjusted SAFT-0 improves the GT and CNT calculations by factors of 500 and 100, respectively. Moreover, it significantly improves the $T$ dependence of GT. Furthermore, both of the CNT and GT have a good $S$ dependence at high temperatures while it is slightly better in CNT at low temperatures.\\

The GT results were compared with the experimental water data using the following Hale's plots: $\log J$ versus $\log J_{scaled}$, with $\Omega=1.45$, and $\log J$ versus $\ln S_{scaled}$, without assuming a value for $\Omega$. The calculations of GT and the measurements by W\"olk and Strey  scale remarkably well. The scaling of the GT rates is most likely due to the combination of a nonclassical nucleation theory that avoids the drastic
and unphysical assumptions of CNT (incompressible bulk liquid
with a sharp vapor-liquid interface~\cite{gibbs_j._willard_scientific_1961,obeidat_nucleation_2004}) and an accurate equation of state (adjusted SAFT-0 EOS) that treats the effects of association. Also, we expect that one can significantly improve the calculations of water nucleation rates by applying gradient theory using our adjusted SAFT-0 EOS and Wilhelmsen et al.~\cite{wilhelmsen2015,wilhelmsen2015a} technique of including the curvature-dependence of surface tension.

We conclude that the adjusted SAFT-0 EOS is satisfactorily exact and predictive, thus future computational studies of more complex systems, such as aqueous-alcohol mixtures, are now expected to be more convenient and more accurate, in particular,  below $T_{max}$. \\

\section*{Acknowledgements}
The author acknowledges Prof. Gerald Wilemski and Prof. Barbara Hale for valuable discussions and Dr. Abdalla Obeidat for sharing his GT code. He also acknowledges F. AlFaran and P. Field from HCT for their linguistic comments.

\section*{References}

\bibliography{SAFTBIB}

\end{document}